# Data Distribution Optimization using Offline Algorithms and a Peer-to-Peer Small Diameter Tree Architecture with Bounded Node Degrees

Mugurel Ionut Andreica*, Eliana-Dina Tirsa*, Nicolae Tapus*

* Computer Science and Engineering Department, Politehnica University of Bucharest, Bucharest, Romania
(e-mail: {mugurel.andreica, eliana.tirsa, nicolae.tapus}@cs.pub.ro)

**Abstract:** Multicast data transfers occur in many distributed systems and applications (e.g. IPTV, Grids, content delivery networks). Because of this, efficient multicast data distribution optimization techniques are required. In the first part of this paper we present a small diameter, bounded degree, collaborative peer-to-peer multicast tree architecture, which supports dynamic node arrivals and departures making local decisions only. The architecture is fault tolerant and, at low arrival and departure rates, converges towards a theoretically optimal structure. In the second part of the paper we consider several offline data distribution optimization problems, for which we present novel and time-efficient algorithmic solutions.

## 1. INTRODUCTION

Multicast data transfers occur frequently in many types of distributed systems, like Grids, IPTV systems, content delivery networks and distributed databases. Although multicast transfers can be performed as multiple unicast streams, this approach is very bandwidth-inefficient. The most common multicast method is to construct, maintain and use a multicast tree. This way we can save bandwidth and reduce overall network traffic. However, this approach poses new challenges, like load balancing (each node should be connected to only a few other nodes), reducing latency and/or hop count (the path between any two nodes in the tree should not be too long) and efficient, scalable construction and maintenance. In the first part of this paper we present a peer-to-peer multicast tree topology with small diameter, where the degree (number of connections) of each node is bounded by a chosen value $K$. The topology can seamlessly handle node arrivals and departures using local decisions only, in a number of time steps proportional to the diameter of the multicast tree. Due to its low diameter, the architecture provides paths with small hop counts from every peer to every other peer. An interesting feature of the architecture is that, at low arrival and departure rates, the topology converges towards a theoretically optimal structure. In the second part of the paper we consider several offline data distribution optimization problems, for which we provide new and efficient algorithms, which improve upon some of the existing solutions in the literature.

The rest of this paper is structured as follows. In Section 2 we describe the bounded degree small diameter multicast tree in detail and analyze it both from a theoretical and a practical point of view. In Section 3 we consider the problem of covering the vertices of a tree by a minimum weight subset of multicast groups (subtrees). A simpler version of this problem has been considered before, but we present an improved dynamic programming algorithm. In Section 4 we present a new solution to a restricted case of the problem of maximum profit scheduling of data transfer requests using conflict graphs. In Section 5 we present new algorithms for asymmetric binary search of unknown parameters, when both costs (durations) and resource consumption are involved. In Section 6 we compute the number of packet permutations with several types of inversions. In Section 7 we discuss related work and in Section 8 we conclude.

## 2. BOUNDED DEGREE SMALL DIAMETER MULTICAST TREE

Maintaining a small-diameter multicast tree over all the peers of a distributed system is a desirable feature in several types of applications. For instance, in Internet TV and live streaming applications, it is more bandwidth-efficient to use a multicast tree instead of sending multiple unicast streams. Moreover, by using a self-organizing multicast tree, there is no need for the content source to be aware of all the peers in the group. Some of these content distribution applications require that the latency of each path from the source to a destination be as small as possible. In this respect, it is desirable for the multicast tree to have a small diameter (diameter=the largest distance between any two nodes in the tree). If the tree has a small diameter, then any of the tree nodes can become a content producer and distribute its content (or send content search queries) to all the other peers in the tree efficiently. Another condition for a good multicast tree is for the traffic load on each node of the tree to be equitably distributed. We can quantify this request in many ways. In this paper we consider a simple measure: the degree of every node in the tree must be bounded from above by a (small) fixed value $K \geq 2$. Although it is possible for every peer to use its own value of $K$, in this paper we will consider only the case when all the peers make use of the same value $K$. In this section we present an implementation of the multicast tree based on a peer-to-peer topology. The topology maintains bounded degrees for all the nodes, has small diameter and supports node arrivals and departures. The neighbouring peers in the topology periodically exchange information among each other (gossip), which is particularly useful when a peer joins or leaves the topology. We will describe next the gossiping, joining and leaving processes for the peer-to-peer architecture.

*2.1 Gossiping in the Multicast Tree*

Periodically, every peer in the tree sends two types of gossiping messages. The first type is sent to the peers at distance (at most) two in the tree and simply broadcasts its existence to these peers. Thus, every peer *X* knows all the peers located at distance one (neighbors) and two (2-neighbors) from *X* in the tree. For every 2-neighbor *Z*, peer *X* maintains the neighbor *Y* which is on the path between *X* and *Z*. Since the degree of every peer is at most *K*, every peer *X* is aware of at most $K+K \cdot (K-1)=K^2$ other peers. Every peer *X* maintains two estimated values for every tree neighbor *Y*: *NumPeers(X,Y)* and *Dmax(X,Y)*. *NumPeers(X,Y)* is an estimate of the total number of peers in *T(X,Y)*=the part of the tree which contains peer *Y* but does not contain peer *X* (i.e. if we consider the tree rooted at *X*, then peer *Y* is a son of peer *X* and *T(X,Y)* is the subtree rooted at peer *Y*); see also Fig. 1. *Dmax(X,Y)* is an estimate of the longest path (in terms of peers) from peer *Y* to the farthest peer in *T(X,Y)*.

The second type of gossiping message is sent by every peer *Y* to every tree neighbor *X* and contains the new values *NumPeers(X,Y)* and *Dmax(X,Y)* that peer *X* should use. Peer *Y* computes *NumPeers(X,Y)* based on its own values *NumPeers(Y,\*)*. Let's denote by *SumNumPeers(Y)* the sum of all the values *NumPeers(Y,\*)* stored by peer *Y*. Then *NumPeers(X,Y)=SumNumPeers(Y)-NumPeers(Y,X)+1*. Let *DistMax(Y, j)* be the $j^{th}$ largest distance among all the values *Dmax(Y,\*)* and let *DistMaxNeigh(Y,j)* be the neighbor *Z* such that *Dmax(Y,Z)=DistMax(Y,j)* and *Z≠DistMaxNeigh(p)* for all *1≤p≤j-1* (*DistMax(Y,j)=0* and *DistMaxNeigh(Y,j)=undefined* if *j* is larger than the number of neighbors peer *Y* has). We will compute *DistMax(Y,j)* and *DistMaxNeigh(Y,j)* only for *j=1,2*. The value *Dmax(X,Y)* sent by peer *Y* to peer *X* is computed as follows: if *DistMaxNeigh(Y,1)≠X*, then *Dmax(X,Y)=1+DistMax(Y,1)*; otherwise, *Dmax(X,Y)= 1+DistMax(Y,2)*. These values (*NumPeers(X,Y)* and *Dmax(X,Y)*) are only estimates of the total number of peers in *T(X,Y)* and of the longest path in *T(X,Y)* starting at *Y*, because they are not immediately updated whenever a new peer joins the system or an old peer leaves the system. However, we will show that, if no peer joins or leaves the system, these values converge to the actual correct values after a number of gossiping periods which is proportional to the diameter of the tree. In order to present the proof, we will define the concept of *layer of leaves*. A leaf in the tree is a vertex with degree *1*. *L(1)* is the set of all the leaf nodes of the tree. *L(i≥2)* is the $i^{th}$ layer of leaves, composed of those nodes which become leaves in the tree if we remove all the nodes in the sets *L(j)* (*1≤j≤i-1*). We assume that the tree has *LL* layers of leaves. It is well-known that the last layer, *L(LL)*, contains only one or two adjacent nodes (the center or the bi-center of the tree); *LL* is equal to *(D+1)/2*, where *D* is the diameter of the tree (length of the longest path in the tree, expressed in terms of tree edges). If *L(LL)* contains two nodes *A* and *B*, we will add an extra layer *LL+1* and move one of the nodes (*A* or *B*) to that extra layer (and then set *LL=LL+1*). Thus, we will consider that *L(LL)* contains only one node.

The values *NumPeers(\*,\*)* and *Dmax(\*,\*)* converge to the corresponding correct values in *O(D)* gossiping periods. We will first show that the values *NumPeers(X,Y)* and *Dmax(X,Y)*, with *X* located on a layer *Q* higher than the layer of *Y*, converge to the correct values in at most *Q-1* gossiping periods. We will prove this by induction on the layer number of the peer *X*. The assumption is true for all the peers *X* in *L(1)*, because they have no neighbor *Y* located on a lower layer. Let's assume now that the proposition is true for all the peers on the layers *1,…,i* and we will prove it for the layer *i+1*. Peer *X* from *L(i+1)* receives the information from a peer *Y* in *L(j)* (*j≤i*). The value *NumPeers(X,Y)* sent by peer *Y* to peer *X* is equal to the sum of the values *NumPeers(Y,W)*, with *W≠X*. Due to the properties of any tree graph, peer *Y* can have only one neighbor on a layer of leaves with an index higher than *j*; this neighbor is *X*. Thus, all the other neighbors *W* are located on layers which are lower than *j* and, by the induction hypothesis, the values *NumPeers(Y,W)* become correct in less than *i* periods. As a consequence, the value *NumPeers(X,Y)* will become correct at the next gossiping period. The same holds for *Dmax(X,Y)*, which is equal to *1+max{Dmax(Y,W)| W≠X is a neighbor of peer Y}*, because the values *Dmax(Y,W)* become correct in at most *i-1* periods. After all the values *NumPeers(X,Y)* and *Dmax(X,Y)* with peer *Y* located on a lower layer of leaves than peer *X* become correct, we will prove that all the values *Dmax(Y,X)* become correct in at most *LL-j* extra periods, where Y belongs to *L(j)*. We will prove this in decreasing order of the index of the layer of leaves of the peer *Y*. For the single peer *A* in *L(LL)* this is true, because it has no neighbors located on a higher layer. Let's assume that the proposition is true for all the peers located on the layers of leaves *LL, LL-1, …, i*. We will now show that the values of all the peers on the layer *i-1* become correct after (at most) *LL-i+1* extra periods. Let's consider a peer *X* from *L(i-1)* and a neighboring peer *Y* from *L(j)* (*j≥i*). Peer *X* receives the values *NumPeers(X,Y)* and *Dmax(X,Y)* from peer *Y*. *NumPeers(X,Y)* (*Dmax(X,Y)*)is computed based on the values *NumPeers(Y,W)* (*Dmax(Y,W)*), with *W≠X*. From the induction hypothesis, all the values *NumPeers(Y,\*)* and *Dmax(Y,\*)* are correct after *LL-i* extra periods. Thus, *NumPeers(X,Y)* and *Dmax(X,Y)* will be correct at the next gossiping period. This concludes our proof.

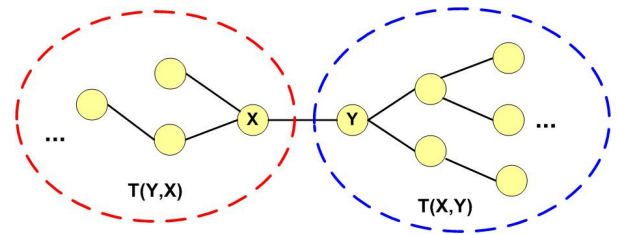

Fig. 1. T(X,Y) and T(Y,X) for 2 neighbouring peers X and Y.

*2.2 Joining the Multicast Tree*

When a new peer *X* wants to join the multicast tree, it must know how to contact any other peer *Y* which is already part of the tree (the peer *Y* can be any peer already in the tree). During the joining procedure, peer *X* will be gradually redirected to other peers until it reaches a peer to which it will connect in the tree. Whenever peer *X* contacts a new peer

$Y$ in order to join the tree, it will also tell peer $Y$ which other peer $Z$ redirected peer $X$ to peer $Y$ (peer $Z$ will be a neighbor of peer $Y$). At the initial join contact, the previous peer $Z$ will be undefined. Let's assume that peer $X$ contacted a peer $Y$ in order to join the system and was redirected here by peer $Z$ (or by nobody if this is the first join attempt, in which case $Z$ is undefined). Peer $Y$ will consider all of its (at most) $K$ tree neighbors $W$, except for $W=Z$. For each neighbor peer $W \neq Z$, peer $Y$ knows the estimates $NumPeers(Y,W)$ and $Dmax(Y,W)$. Then, peer $Y$ computes the maximum number of peers $MaxNumPeers(Y,W)$ which can be located in $T(Y,W)$ such that $Dmax(Y,W)$ does not increase. $MaxNumPeers(Y,W)$ is equal to $1+(K-1)+(K-1)^2+\ldots+(K-1)^{Dmax(Y,W)-1}$. If $K=2$, then $MaxNumPeers(Y,W)=Dmax(Y,W)$; else, $MaxNumPeers(Y,W)=((K-1)^{Dmax(Y,W)}-1)/(K-2)$. If $MaxNumPeers(Y,W) > NumPeers(Y,W)$, then peer $W$ is a *valid* neighbor; otherwise, it is not valid. Among all of peer $Y$'s valid neighbors $W \neq Z$, peer $Y$ will choose the peer $W_{next}$ as the one with the smallest value $Dmax(Y,W_{next})$ (if there are multiple such neighbors, one will be chosen arbitrarily). Peer $X$ will be redirected to the peer $W_{next}$. If peer $Y$ has no valid neighbors and peer $Y$'s degree is less than $K$, then peer $Y$ will connect directly to peer $X$. Peer $X$'s degree will now be $1$ (it will be a leaf in the tree) and peer $Y$'s degree increases by $1$. $NumPeers(Y,X)$ and $Dmax(Y,X)$ will be $1$; $Dmax(X,Y)$ and $NumPeers(X,Y)$ will be sent immediately to peer $X$ (they will be computed as described in subsection 2.1). If peer $Y$ has no valid neighbors and its degree is equal to $K$, then it will choose the neighbor $W_{next} \neq Z$ with the smallest value $Dmax(Y,W_{next})$ (disregarding the values $NumPeers(Y,W_{next})$ and $MaxNumPeers(Y,W_{next})$). Peer $X$ will be redirected to peer $W_{next}$. If peer $X$ was redirected to another peer $W_{next}$, at the next join request peer $X$ will contact peer $W_{next}$ and will tell it that it was redirected there from peer $Y$. We can see that peer $X$ may be redirected (at most) a number of times proportional to the diameter of the tree.

*2.3 Leaving the Multicast Tree*

When a peer $X$ leaves the multicast tree (gracefully or suddenly), its tree neighbors will detect this event (because every neighbor will periodically send *keep-alive* and *ping* messages to both its neighbors and its 2-neighbors). Because of the first type of gossiping messages, every neighbor $Y$ of peer $X$ knows every other neighbor of peer $X$. Every neighbor $Y$ will compute the value $DistMaxNoX(Y,X)$; if $DistMaxNeigh(Y,1)=X$, then $DistMaxNoX(Y,X)=DistMax(Y,2)$; otherwise, $DistMaxNoX(Y,X)=DistMax(Y,1)$. Each (former) neighbor $Y$ of peer $X$ will send the value $DistMaxNoX(Y,X)$ to every other (former) neighbor of $X$, as well as a unique, self-generated identifier (e.g. the result of a hash function). The (former) neighbor $W$ of peer $X$ with the largest value $DistMaxNoX(W,X)$ will be chosen by every other (former) neighbor as their representative (if multiple neighbors $Z$ have the same largest $DistMaxNoX(Z,X)$ value, ties will be broken by considering the unique identifiers of the peers; e.g. the peer with the smallest or largest identifier will be chosen). From a practical point of view, each (former) neighbor $Y$ of peer $X$ will wait at most a certain amount of time for receiving the corresponding values from any other (former) neighbor $Y'$ of peer $X$. Since 2-neighbors periodically ping each other, peer $Y$ can have a good estimate of the latency $lat$ of the network path to any 2-neighbor $Y'$; peer $Y$ can wait for the information from $Y'$ for at most $C \cdot lat$ time units, where $C \geq 2$ is a constant.

Peer $W$ will send a message to the peer $Q$ for which the path from $W$ to $Q$ in the tree contains exactly $DistMax(W,1)$ peers (if $DistMax(W,1)$ has converged to the correct value). Peer $W$ does not need to know peer $Q$ before-hand. Peer $W$ will forward the message to its neighbor $W' \neq X$ with the largest value $Dmax(W,W')$. Whenever a peer $W'$ receives the message from a peer $W''$, it will forward it to the neighbor $W''' \neq W''$ with the largest value $Dmax(W', W''')$. Note that the neighbor $A \neq V$ of a peer $B$ with the largest value $Dmax(B,A)$ can be computed in $O(1)$ time: if $DistMaxNeigh(B,1) \neq V$, then $A=DistMaxNeigh(B,1)$; otherwise, $A=DistMaxNeigh(B,2)$. Eventually, the message will reach a peer $Q$ which is a leaf in the tree and, thus, cannot forward the message further. If all the values $Dmax(*,*)$ have converged to their stable states, then the path from peer $W$ to peer $Q$ is the longest path from peer $W$ to any peer in its part of the tree ($T(X,W)$); otherwise, this path is only an approximation of the actual longest path (although we may obtain the longest path even if the $Dmax(*,*)$ values have not converged, yet). Peer $Q$ will disconnect from its only neighbor in the tree (if the representative peer $W$ had no other neighbors except peer $X$, then $Q=W$ and no disconnection is performed) and will replace peer $X$; that is, peer $Q$ will connect to all the former neighbors of peer $X$. Thus, after a peer $X$ departs from the tree, the tree returns to a correct structure after a number of time steps which is proportional to $K+D$, where $D$ is the diameter of the tree. After connecting to all the former neighbors $Y$ of peer $X$, peer $Q$ receives the values $NumPeers(Q,Y)$ and $Dmax(Q,Y)$ from these neighbors. As soon as it receives all of these values, peer $Q$ will send back the values $NumPeers(Y,Q)$ and $Dmax(Y,Q)$ to every neighbor $Y$ (all these values are computed the way we showed in a previous subsection).

In order to minimize the period of time during which the tree remains disconnected after the departure of a peer $X$, we can use a proactive approach, instead of the reactive approach presented above. Every peer $Y$ periodically computes the values $DistMaxNoX(Y,Z)$ (as described previously) and $Q_{far}(Y,Z)$=the peer $Q$ which would be chosen by the previously described method, if the neighbor $Z$ of peer $Y$ were to leave the tree. Thus, if peer $Y$ is chosen as the representative peer among all the neighbors of a departed peer $X$, then the peer $Q$ which will replace $X$ is $Q_{far}(Y,X)$. Moreover, every 2 neighbors $Y$ and $Z$ of a peer $X$ could periodically exchange between them the values $DistMaxNoX(Y,X)$ and $DistMaxNoX(Z,X)$ (together with their identifiers). This way, when a peer $X$ leaves the tree, every former neighbor $Y$ of peer $X$ already knows the values $DistMaxNoX(Z,X)$ of all the other former neighbors $Z$ of peer $X$ and can immediately select the representative (former) neighbor $W$. With this proactive approach, the tree stays disconnected only for a very short time ($O(1)$ time steps) whenever a peer $X$ leaves the tree.

*2.4 Experimental Tests*

In order to test the multicast tree peer-to-peer topology, we developed a simulation framework, which we implemented in the Python programming language. We performed two types of tests. The first tests were *incremental* tests. *600* peers were added sequentially, at different rates, and considering two values of *K* (*3* and *6*). The rate was measured as the number of newly added peers divided by the number of gossiping periods. We measured the tree's diameter after every peer addition. The lowest rate was *1/D*, where *D* was the (current) diameter of the tree; obviously, this rate was not constant. At this rate, all the *NumPeers(\*,\*)* and *Dmax(\*,\*)* values became correct before the next peer addition. The consequence was that the diameter of the obtained tree was always equal to the theoretical optimum (i.e. the diameter of a perfectly balanced tree with the same number of nodes as the multicast tree and with the same upper bound on the node degrees). We considered both the case when every peer started its joining process from a random peer and the case when all the peers started from the same (first) peer. The same cases were considered for other rates: *2/D*, *1*, *2* and *5*. As expected, the higher the rate, the higher the tree's diameter was (however, there was no difference in the diameters for the rates *1/D* and *2/D*). Fig. 2 (right) presents the diameters obtained for *K=3* and different ratios, as a function of the number of peers, when every peer joined the tree starting from another random peer. The results for the case when every peer started the joining process from the same peer are similar. Fig. 2 (left) shows the obtained tree topology for *K=3* and *100* peers. The tests of the second type were *decremental*. We started from the tree with *600* peers and optimal diameter and repeatedly removed from the tree the peer X whose largest estimate *Dmax(X,\*)* was minimum (i.e. the tree's center). The tree recovered gracefully every time and maintained the optimal theoretical diameter after every peer removal. A peer was removed only after the tree recovered correctly from the previous peer removal.

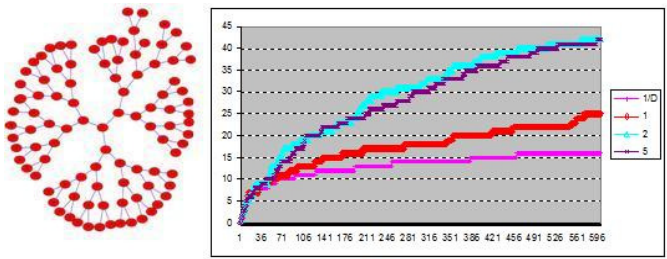

Fig. 2. Left - Multicast Tree with 100 peers (K=3). Right - Tree diameter after every peer insertion at different peer insertion ratios.

## 3. COVERING THE VERTICES OF A TREE BY MULTICAST GROUPS

We consider a tree with *n* vertices and *m* multicast groups. Each group *j* (*1≤j≤m*) is a subtree of the tree and has a weight *w(j)≥0*. We want to find a subset of groups whose total weight is minimum, such that every tree vertex belongs to at least one group in the subset. This problem is motivated by the need to broadcast important information to all the nodes of a network with a tree topology, when the nodes are part of several multicast groups and all we can do is broadcast the information within some of the multicast groups. We will present a dynamic programming solution for the case when every vertex *i* belongs to a bounded number of groups *C*. A simpler version of this problem (where every group is a path) has been considered in (Lin et al., 2006), where an $O(m+n^2 \cdot 2^{2 \cdot C})$ solution was given. Our algorithm has a better time complexity ($O(m+n \cdot 2^C)$) and handles a more general case.

We consider the tree rooted at an arbitrary vertex *r*. In the rooted tree we define *parent(i)*=vertex *i*'s parent, *T(i)*=vertex *i*'s subtree, *ns(i)*=the number of sons of vertex *i*, and *s(i,1), ..., s(i,ns(i))*, the sons of vertex *i*. For each vertex *i* we denote by *ng(i)* the number of groups into which vertex *i* is included. The identifiers of the groups which contain vertex *i* are *g(i,1), ..., g(i,ng(i))* (*1≤g(i,j)≤m*, *1≤j≤ng(i)*). We denote by *ncg(i)* (*0≤ncg(i)≤ng(i)*) the number of groups containing vertex *i* which also contain *parent(i)* (*ncg(r)=0*). For each vertex *i*, we sort the group identifiers in such a way that *g(i,1), ..., g(i,ncg(i))* are the groups containing both *i* and *parent(i)*. For each vertex *i* we will consider $2^{ng(i)}$ states. A state *S=(sel(1), ..., sel(ng(i)))* (*sel(j)=0,1*) has the following meaning: if *sel(j)=1*, then the group *g(i,j)* is chosen to be part of the solution; otherwise, *g(i,j)* is not part of the solution. We also denote the $j^{th}$ component of the state *S* by *S(j)*. We denote by *pg(i,gj)* the position where the group numbered *gj* appears in the list *g(i,\*)*, i.e. we have that *g(i,pg(i,gj))=gj*. For each *(vertex, state)* pair *(i,S)*, we will compute $W_{min}(i,S)$=the minimum total weight of a subset of groups covering all the vertices in *T(i)*, with the property that the states of the groups containing vertex *i* are given by the state *S*. We will compute these values bottom-up. Let *wSum(i,S)* be:

$$wSum(i,S) = \sum_{j=1}^{ng(i)} S(j) \cdot w(g(i,j)) \quad (1)$$

We have that $W_{min}(*,(0, ..., 0))=+\infty$. For a leaf vertex *i*, $W_{min}(i, S\neq(0, ..., 0))=wSum(i,S)$. After computing the values $W_{min}$ for every pair *(i,S)* of any vertex *i* (leaf or non-leaf), we will compute some values $W'_{min}(i,S')$ for every state *S'=(sel(1), ..., sel(ncg(i)))*. Basically, we initialize $W'_{min}(i,S')=+\infty$ (for all the states *S'*) and then, for every state *S=(sel(1), ..., sel(ng(i)))*, we set $W'_{min}(i,(sel(1), ..., sel(ncg(i))))=\min\{W'_{min}(i,(sel(1), ..., sel(ncg(i)))), W_{min}(i,S)\}$. We will also compute

$$wSum'(i,S') = \sum_{j=1}^{ncg(i)} S'(j) \cdot w(g(i,j)) \quad (2)$$

For a non-leaf vertex *i* and a state *S=(sel(1), ..., sel(ng(i)))≠(0, ..., 0)*, we have:

$$W_{min}(i,S) = wSum(i,S) + \sum_{p=1}^{ns(i)} \begin{pmatrix} W'_{min}(s(i,p), S' = (S(pg(i,g(s(i,p),1))),..., \\ S(pg(i,g(s(i,p),ncg(s(i,p)))))) \\ - wSum'(s(i,p),S') \end{pmatrix} \quad (3)$$

We can easily implement this algorithm in $O(m+n \cdot C \cdot 2^C)$ time. If we compute the sums for every state more carefully, we obtain an $O(m+n \cdot 2^C)$ complexity. We will maintain the current sum (initially *0*) and the current state as two global variables *sum* and *S* during the state generation via backtracking. The initial call should be *computeSums(1, i)*.

**computeSums(level,i):**
**if** *(level>ng(i))* **then** { // or (level>ncg(i))
 *wSum(i,S)=sum* // or wSum'(i,S)=sum }

```
else {
  S(level)=1; sum=sum+w(g(i,level));
  computeSums(level+1,i)
  S(level)=0; sum=sum-w(g(i,level))
  computeSums(level+1,i) }
```

The optimal cost is $min\{W_{min}(r,*)\}$ and the subset of selected groups can be computed easily from the $W_{min}(*,*)$ values. If, instead of covering all the tree vertices we want to cover all the tree edges (which also implies covering all the vertices), we can modify the dynamic programming algorithm as follows. For every non-leaf vertex $i$ and every state $S$ such that there exists at least one son $s(i,j)$ for which $S(pg(i,g(s(i,j),k)))=0$ (for all $1 \leq k \leq ncg(s(i,j))$), we will consider $W_{min}(i,S)=+\infty$. To be more precise, for every vertex $i$, we cannot consider states in which all the groups containing both $i$ and a son $s(i,j)$ of $i$ are not selected to be part of the solution.

## 4. MAXIMUM PROFIT SCHEDULING OF DATA TRANSFER REQUESTS USING CONFLICT GRAPHS

We consider here the following scheduling problem. We have $N$ (multicast or point-to-point) data transfer scheduling requests. The time horizon over which the data transfers can be scheduled is divided into $T$ time slots (numbered from $1$ to $T$). A request $i$ asks for exclusive access to a specific set of network links during a given interval of time slots $[ts(i), tf(i)]$ and, if accepted, it brings a profit of $p(i)$. Two requests whose time slot intervals overlap may be in conflict if they ask for exclusive access to at least one common network link. We will model these conflicts by using a conflict graph $CG$ in which we have a vertex for every request $i$ ($1 \leq i \leq N$) and an edge between two requests $i$ and $j$ if their intervals overlap and they are in conflict. Using this model, we want to find an independent set $IS$ within the conflict graph (i.e. there is no edge between any two requests $a$ and $b$ from $IS$), such that the sum of the profits of the requests in $IS$ is maximum. All the requests in $IS$ will be accepted and all the other requests will be rejected. The problem of computing a maximum weight independent set in an arbitrary graph is an NP-hard problem. We will present here a solution for a restricted case. We maintain two lists of events for each time slot $t$: a list $LAE(t)$ with *activation events* (when a new request becomes active) and a list $LDE(t)$ with *deactivation events* (when a request becomes inactive). An activation event for a request $i$ is added to the list $LAE(ts(i))$ and the deactivation event is added to the list $LDE(tf(i)+1)$. We will traverse the time slots in increasing order and, during the traversal, we will maintain a set $S$ of subsets of requests: $S(0), ..., S(k-1)$ ($k=|S|$). $S(0)$ will always exist and will always be void (empty). For each time slot $t$ ($1 \leq t \leq T$), in increasing order, we will handle all the events in $LDE(t)$ first, followed by all the events in $LAE(t)$. For each deactivation event for a request $i$, we find the set $S(j)$ which contains the request $i$ and remove $i$ from $S(j)$; if $S(j)$ becomes void, we will remove $S(j)$ from $S$. For each activation event for a request $i$, we will consider all the requests $j$ such that there exists an edge $(i,j)$ in the conflict graph. Let $S(jj(1)), ..., S(jj(q))$ be the subsets which contain all the requests $j$ which are $i$'s neighbors in $CG$ (some of these neighbors may not belong to any subset $S(x)$, because their activation events have not been handled, yet; these neighbors will be ignored). We will construct a set $SQ$ from the union of $S(jj(1)), ..., S(jj(q))$, and then remove all these subsets from $S$. Afterwards, we will add $i$ to $SQ$ and we will insert $SQ$ into $S$. We will present an algorithm which will use the subsets $S(*)$ and which is efficient in the following case: at any moment during the execution of the algorithm, the cardinality $|S(j)|$ of any subset $S(j)$ is at most $CMAX$, where $CMAX$ is a small constant value (i.e. the cardinality is bounded by a constant).

Let the vertices of a subset $S(j)$ be $v(j,1), ..., v(j,|S(j)|)$. For each subset $S(j)$ we will maintain a table $T_j(state)$, where *state* is a tuple with $|S(j)|$ binary values (i.e. $0$ or $1$); we denote the $i^{th}$ of these values by $state(i)$. If $state(i)=1$, then we consider that $v(j,i)$ belongs to $IS$; otherwise, $v(j,i)$ does not belong to $IS$. There are $2^{|S(j)|}$ such states. Every vertex $x$ of $CG$ can be assigned a label $label(x)=p$ which means that the activation event of vertex $x$ was/will be the $p^{th}$ such event processed during the algorithm ($p \geq 1$). The value $T_j(state)$ is equal to the maximum profit which can be achieved if the vertices $v(j,*)$ are in the state defined by *state*, and we have already considered all the vertices $x$ with $label(x) < min\{label(v(j,*))|1 \leq j \leq |S(j)|\}$ which are in the same connected component of $CG$ as the vertices $v(j,*)$. If $S(j)$ contains no vertices, then we have only one possible state, which is the empty tuple *{}*. If we traverse the time slots all the way to $T+1$, then we will eventually process all the deactivation events and, in the end, the only remaining subset in $S$ will be the empty set, $S(0)$. $T_0(\{\})$ will be the maximum weight of an independent set $IS$ of $CG$. We will now describe how the values $T_j(*)$ are maintained by the algorithm after processing every activation and deactivation event. Initially, we only have the set $S(0)$, with $T_0(\{\})=0$. When the algorithm processes the deactivation event of a request $i$, it finds the set $S(j)$ which contains the request $i$. Let's assume, w.l.o.g., that, within the set $S(j)$, we have $v(j,|S(j)|)=i$ (we can change the order of the vertices in $S(j)$ such that $i$ is the last vertex). We will now consider every state $s$ with $|S(j)|-1$ binary values and we will compute a new table $T_{new,j}(*)$, where $T_{new,j}(s)= max\{T_j(s(0), ..., s(|S(j)|-1), 0), T_j(s(0), ..., s(|S(j)|-1), 1)\}$. Afterwards, we remove vertex $i$ from $S(j)$ and we replace $T_j$ by $T_{new,j}$ (i.e. we set $T_j=T_{new,j}$) If $S(j)$ contains no more vertices, then we add $T_j(\{\})$ to $T_0(\{\})$ and, afterwards, we remove $S(j)$ from $S$. When the algorithm processes an activation event for a request $i$, it proceeds as follows. It finds the sets $S(jj(1)), ..., S(jj(q))$ which contain all the neighbors $j$ of $i$ with $label(j)<label(i)$. Then, it constructs the set $SQ$ as the union of these sets. We will consider that the vertices of $SQ$ are ordered as follows: $v(jj(1),1), ..., v(jj(1),|S(jj(1))|), v(jj(2), 1), ..., v(jj(2), |S(jj(2))|), ..., v(jj(q),1), ..., v(jj(q), |S(jj(q))|)$. Then, we will compute a table *Taux*. We consider every combination of states $st(jj(1)), ..., st(jj(q))$, corresponding to the sets $S(jj(1)), ..., S(jj(q))$ and we set $Taux(st(jj(1),1), ..., st(jj(1), |S(jj(1))|), st(jj(2), 1), ..., st(jj(2), |S(jj(2))|), ..., st(jj(q), 1), ..., st(jj(q), |S(jj(q))|))= T_{jj(1)}(st(jj(1)))+...+ T_{jj(q)}(st(jj(q)))$. If the set $SQ$ is empty, then the table *Taux* contains only one entry, corresponding to the empty tuple: $Taux(\{\})=0$. Afterwards, we will construct the set $SQ'$, as the union of $SQ$ and $\{i\}$ ($i$ will be the last vertex in $SQ'$) and we will compute a table $Taux_2$. For every state $stq$ for which an entry exists in *Taux*, we set $Taux_2(stq(1), ...,$

$stq(|SQ|), 0)=Taux(stq)$. Then, for each such state $stq$, let's consider the positions $pos$ corresponding to the neighbors $j$ of $i$ with $label(j)<label(i)$. If $stq(pos)=0$ for every such position, we set $Taux_2(stq(1), …, stq(|SQ|), 1)=Taux(stq)+p(i)$; otherwise, we set $Taux_2(stq(1), …, stq(|SQ|), 1)=-\infty$. Then, we add the set $SQ'$ to $S$ (after removing from $S$ all the sets $S(jj(1)), …, S(jj(q))$). If $SQ'$ is assigned the index $p$ (i.e. $SQ'=S(p)$), then we set $T_p=Taux_2$. The time complexity of the algorithm is $O(T+N \cdot 2^{CMAX})$. By maintaining the tables $T_j(*)$ after each processed event, we can compute the actual solution (which requests are accepted) and not just the maximum profit. We mention that the algorithm also works without dividing the time into time slots. In this case, every request has a time interval $[ts(i), tf(i)]$ and we construct an activation event $(time=ts(i), type=+1, request=i)$ and a deactivation event $(time=tf(i), type=-1, request=i)$. We then sort these requests, first in increasing order of the time moment and, for equal time moments, we place the deactivation events before the activation events occurring at the same time moment. In this case, the complexity is $O(N \cdot log(N)+N \cdot 2^{CMAX})$.

## 5. MINIMUM COST BINARY SEARCH STRATEGY FOR DETERMINING UNKNOWN PARAMETERS

In this section we present novel algorithmic solutions for determining optimal binary search strategies, when costs and resources are involved.

### 5.1 Minimizing the Number of Tests

We have an unknown parameter $F$ which is an integer value between $0$ and $N$. In order to guess the value $F$, we have a procedure $Test(x)$, which returns *true*, if $F \geq x$, and *false*, if $F<x$. Obviously, we have $1 \leq x \leq N$ (as $Test(0)$ always returns *true*). We have $M$ units of a resource, which is used during testing. If the answer of a test $Test(x)$ is *false*, then $1$ unit of the resource is consumed; otherwise, zero units are consumed. In order to perform a test, we need at least $1$ unit of this resource. We want to devise a strategy which performs the minimum number of tests, in the worst case. If $M=1$, we are forced to perform the tests $Test(1), Test(2), …, Test(k)$, until $Test(k)=false$ or $k=N$. If $k=N$ and $Test(N)=true$, then $F=N$; otherwise, $F=k-1$. Thus, we need $N$ tests in the worst case. Intuitively, we can do better for $M \geq 2$. We will compute a table $T(i,j,k)$=the minimum number of tests required in the worst case, if we have $k$ resource units available and the value of $F$ is between $i-1$ and $j$. $T(1,N,M)$ is the answer to our problem. If $i=j+1$ then $T(i,j,k)=0$ (because $F=j$). If $k=1$, $T(i,j,1)=j-i+1$. For $k \geq 2$, we will compute the table in ascending order of the value $(j-i)$. For $i<j$, we will consider every possible value $p$ between $i$ and $j$ for which we could perform the next test ($Test(p)$). Let's assume that we perform the test $Test(p)$. If the answer is *false*, then $F$ is within the interval $[i-1,p-1]$ and we have $k-1$ resource units left. If the answer is *true*, $F$ will be within the interval $[p,j]$ and we will have $k$ resource units left. In the worst-case, after performing the test with value $p$, we will need to perform $Q(i,j,k,p)=max\{T(i,p-1,k-1), T(p+1,j,k)\}$ extra tests. Thus, $T(i,j,k)=1+min\{Q(i,j,k,p)|i \leq p \leq j\}$. The time complexity of this algorithm is $O(N^3 \cdot M)$. We will present several improvements, until its time complexity becomes $O(N \cdot min\{M, log(N)\})$. At first, we notice that $T(i,j,k)=T(1,j-i+1,k)$, i.e. the exact interval of values is irrelevant for computing the worst case number of tests of the optimal strategy. Thus, we will define $T'(l,k)$=the minimum number of tests required in the worst case if we have $k$ resource units available and the value of $F$ belongs to an interval with $l+1$ elements. We have $T'(0,k)=0$ and $T'(l,1)=l$. We also need to define $Q'(l,k,p)=max\{T'(p-1,k-1), T'(l-p,k)\}$ and we have $T'(l,k \geq 2)=1+min\{Q'(l,k,p)| 1 \leq p \leq l\}$. The second observation is that if $M \geq log(N)$, then we can use binary search in order to find the value of $F$. Thus, for $M \geq log(N)$, the optimal strategy cannot perform more than $log(N)$ tests. As such, we never need more than $log(N)$ resource units. So far, the time complexity was brought down to $O(N^2 \cdot min(M, log(N)))$. In order to reduce it even further, we need to make another observation. Let's denote by $P_{opt}'(l,k)$ the optimal value of $p$ which gives the optimal value $Q'(l,k,p)$ which is used for computing $T'(l,k)$. We have that $P_{opt}'(l+1,k)$ is equal either to $P_{opt}'(l,k)$ or to $P_{opt}'(l,k)+1$. This is because the initial table $T(i,j,k)$ has the *monotonicity property*, i.e. $P_{opt}(i,j,k)$ (defined as the value of $p$ for which $Q(i,j,k,p)$ minimizes the value of $T(i,j,k)$) is located between $P_{opt}(i,j-1,k)$ and $P_{opt}(i+1,j,k)$. From a previous observation, it is easy to conclude that $P_{opt}(i,j-1,k)+1=P_{opt}(i+1,j,k)$, which is equivalent to our observation. With this final observation, we can compute $T'(l,k)$ in $O(1)$ time for every pair $(l,k)$, obtaining the promised $O(N \cdot min\{M, log(N)\})$ time complexity.

### 5.2 Minimum Total Duration of the Tests

We now consider the same problem of finding the unknown parameter value $F$. When we perform a test $Test(i)$, the result is available only after $t(i)$ seconds. This time, we want to minimize the total duration of all the tests (in the worst-case), instead of minimizing their number. We can compute a similar matrix $T(i,j,k)$=the minimum total duration of the tests (in the worst case) if we have $k$ resource units available and the value of $F$ is between $i-1$ and $j$. Again, $T(1,N,M)$ is the answer to the problem. $T(i,j,1)=t(i)+t(i+1)+…+t(j)$ (because the only feasible strategy is to perform the tests $Test(x)$, for $x=i,…j$, until we get a negative answer, or until $x$ exceeds $j$. We also have $T(j+1,j,k)=0$. In order to compute $T(i,j,k)$ ($i \leq j$; $k \geq 2$), we consider every candidate value $p$ for performing the test $Test(p)$ ($i \leq p \leq j$). Let $Q(i,j,k,p)$ be the worst case total duration of the tests, if we perform the test $Test(p)$: $Q(i,j,k,p)=t(p)+max\{T(i,p-1,k-1), T(p+1,j,k)\}$. Then, $T(i,j,k)=min\{Q(i,j,k,p)|i \leq p \leq j\}$. Note how this problem is identical to the previous one, if $t(p)=1$ (for every $1 \leq p \leq n$). The studied problem is a particular version of searching for a value $x$ in a sorted array $a$ with $n$ elements, given a finite (integer) amount of resources $r$. Comparing $x$ against the element $a(p)$ takes $tlow(p)$ time and consumes $clow(p)$ resources if $x<a(p)$, $teq(p)$ time and $ceq(p)$ resources if $x=a(p)$, and $thigh(p)$ time and $chigh(p)$ resources if $x>a(p)$. The optimal searching strategy (which minimizes the total time in the worst case) can be found by adapting slightly the dynamic programming algorithm presented earlier. We compute $T(i,j,r)$=the minimum total time spent in the worst case, if $r$ resource units are available and the searched value lies in the interval $[i,j]$. For $i=j$ and $r \geq max\{clow(i), ceq(i), chigh(i)\}$, we have $T(i,i,r)=max\{tlow(i), teq(i), thigh(i)\}$ (because we are not sure that the searched element is actually equal to the last remaining element in the search interval); for $r<max\{clow(i), ceq(i), chigh(i)\}$, $T(i,j,r)=+\infty$. If we can be sure that the searched element is part of the array, then $T(i,i,r\geq 0)=0$ (because the last comparison is not necessary). For $i>j$ we consider $T(i,j,*)=0$. For $i<j$ and an amount of resources $r$, we need to consider every position $p$ ($i\leq p\leq j$) as a candidate for the next comparison. If we compare the searched value against $a(p)$, then the total amount of time required in the worst case is $Q(i,j,r,p)=max\{tlow(p)+T(i,p-1,r-clow(p)), teq(p), thigh(p)+T(p+1,j,r-chigh(p))\}$. $T(i,j,r\geq 0)=min\{+\infty, min\{Q(i,j,r,p)|i\leq p\leq j, r\geq max\{clow(p), ceq(p), chigh(p)\}\}\}$. The time complexity of this algorithm is $O(n^3)$. Standard search strategies do not consider the consumption of resources and, thus, they have $clow(*)=ceq(*)=chigh(*)=r=0$. When only the number of comparisons is of interest, we set $tlow(*)=teq(*)=thigh(*)=1$. When the equality outcome is not possible, we just set $teq(*)=ceq(*)=0$. In the problem from the previous subsection, equality is not a possible outcome, only the number of comparisons (tests) is of interest, and we have $clow(*)=0$ and $chigh(*)=1$. When the costs and times do not depend on the position $p$ (i.e. they are identical for every position $p$), then we have $T(i,j,r)=T(1,j-i+1,r)$ and, thus, we can reduce the complexity to $O(n^2)$.

## 6. COUNTING PACKET PERMUTATIONS WITH INVERSION PROPERTIES

A source node needs to send a communication flow composed of $n$ packets to a destination node. However, in several situations, when the packets are routed along multiple paths, they may reach the destination in a different order than their intended logical order. In order to compute the probabilities of such occurrences, it is often useful to be able to compute the number of permutations with several types of restricted *inversion properties* (e.g. descent set or number of inversions).

### 6.1 Permutations with a Given Descent Set

Let's consider a permutation with $n$ elements, $pe(1), …, pe(n)$. The descent set of the permutation is the set $\{i|1\leq i\leq n-1, pe(i)>pe(i+1)\}$. We are interested in computing the number of permutations with a given descent set. We will first focus on "zig-zag" permutations, i.e. those permutations whose descent set consists of all the even (or all the odd) numbers in the set $\{1, …, n-1\}$. We can easily compute the number of "zig-zag" permutations with $n$ elements in the following way. First, we will compute all the values $C(i,j)$ ($1\leq j\leq i\leq n$), representing the number of ways of choosing $j$ elements out of a set of $i$ elements. This step takes $O(n^2)$ time overall, by using a well-known formula: $C(i,j)=C(i-1,j-1)+C(i-1,j)$. Then, for each $1\leq i\leq n$, we will compute $P(i)$, the number of zig-zag permutations with $i$ elements. Obviously, $P(0)=P(1)=1$. For $i>1$, we will consider all the even positions $j$ where element $i$ can be placed in the permutation. This leaves $j-1$ positions to the left and $i-j$ positions to the right. There are $C(i-1,j-1)$ ways of selecting the elements on the left and $P(j-1)$ ways of placing them into the $j-1$ positions. The elements on the right are the remaining elements and there are $P(i-j)$ ways of permuting them. Thus, $P(i)$ is equal to the sum of the values $C(i-1,2\cdot k-1)\cdot P(2\cdot k-1)\cdot P(i-2\cdot k)$ ($1\leq k\leq i/2$). The time complexity of the algorithm is $O(n^2)$, multiplied by the complexity of performing arithmetic operations on the numbers $C(*,*)$ and $P(*)$ (which can be $O(n)$ for large numbers, or $O(1)$, if we perform all the operations modulo a small number $M$). We will now consider the case where an arbitrary descent set $D$ is given. The brute force solution is to consider all the $n!$ permutations, compute their descent sets and increment a counter every time a permutation with descent set $D$ is found. A second solution consists of noticing that there are $2^{n-1}$ possible descent sets for a permutation with $n$ elements. Thus, we will compute $P(i,D)$, the number of permutations with $i$ elements and descent set $D$. We have $P(1,\{\})=1$. For each pair $(i\geq 2,D)$, we first initialize $P(i,D)$ to $0$. Then, we consider every pair $(i-1,D)$ and every possible position $j$ ($1\leq j\leq i$) where element $i$ can be inserted into an $(i-1)$-element permutation. If $j=i$, then we set $P(i,D)=P(i,D)+P(i-1,D)$. Otherwise, let $D'=\{q+1|q\in D$ and $q\geq j\} \cup (D\setminus\{j-1, j, j+1,…,i-2\}) \cup \{j\}$; we set $P(i,D')=P(i,D')+P(i-1,D)$. The total number of $P(*,*)$ values is $O(2^1+2^2+…+2^n)=O(2^{n+1})$. The overall time complexity of this algorithm is $O(n\cdot 2^n)$. We will now present a polynomial time algorithm for this problem. We will denote by $D|_i$ the subset of $D$ from which we remove every element larger than or equal to $i$, i.e. $D\setminus\{i,i+1,…,n\}$. We will compute the values $P(i,j)$=the number of permutations with $i$ elements, whose last element is $j$ ($1\leq j\leq i$) and whose descent set is $D|_i$. We have $P(1,1)=1$. For $i>1$ we have: if $((i-1)\in D)$, then $P(i,j)$ is equal to the sum of the values $P(i-1,k)$ ($j\leq k\leq i-1$); if $((i-1) \notin D)$, then $P(i,j)$ is equal to the sum of the values $P(i-1,k)$ ($1\leq k\leq j-1$). A straight-forward implementation of these equations leads to an $O(n^3)$ time algorithm. We can improve the complexity to $O(n^2)$, as follows. After computing all the values $P(i,*)$, we compute the "prefix sums" $SP(i,*)$, where $SP(i,0)=0$ and $SP(i,j>0)=SP(i,j-1)+P(i,j)$. With these values, the equations become: $P(i,j)=SP(i-1,i-1)-SP(i-1,j-1)$, if $((i-1)\in D)$, and $P(i,j)=SP(i-1,j-1)$, if $((i-1)\notin D)$. If we do not perform all the operations modulo a small number $M$ (with $O(1)$ digits), then the numbers $P(*,*)$ may have $O(n)$ digits, and the complexities need to be multiplied by an $O(n)$ factor.

### 6.2 Permutations with k Inversions

An inversion of a permutation $p$ is a pair $(i,j)$, such that $i<j$ and $p(i)>p(j)$. We are interested in computing the number of $n$-element permutations having exactly $k$ inversions ($0\leq k\leq n\cdot(n-1)/2$). We will start with a well-known recursive solution. We will compute $P(i,j)$, the number of permutations with $i$ elements and $j$ inversions. We have $P(1,0)=1$ and $P(i,j>i\cdot(i-1)/2)=0$. For $i>1$ (and every value of $j$), we will iterate over the positions on which we can place the element $i$. If we place $i$ on position $p$ ($1\leq p\leq i$), it will introduce $i-p$ extra inversions in the $(i-1)$-element permutation obtained by removing element $i$. Thus, $P(i,j)$ is equal to the sum of the values $P(i-1,j-(i-p))$ ($max\{0,i-j\}\leq p\leq i$). A straight-forward implementation of these equations leads to an $O(n^2\cdot k\cdot Op(n))$ algorithm, where $Op(n)$ is the complexity of performing arithmetic operations on numbers with $O(n)$ digits (normally, $Op(n)=O(n)$; if we perform the operations modulo a number $M$ with $O(1)$ digits, then $Op(n)=O(1)$). Since $k=O(n^2)$, the algorithm takes $O(n^4 \cdot Op(n))$ time. By using the "prefix sums" technique we mentioned in a previous subsection, we can reduce the time complexity to $O(n \cdot k \cdot Op(n))$. We will compute $SP(i,j)=P(i,0)+\ldots+P(i,j)$. We have $SP(i,-1)=0$ and $SP(i,j \geq 0)=SP(i,j-1)+P(i,j)$. Now we have $P(i,j)=SP(i-1,j)-SP(i-1,\max\{-1,j-i-1\})$. We will now present an improved solution for the case when $n$ is (significantly) larger than $k$. We notice that for $i \geq j$, $P(i,j)$ is equal to the sum of the values $P(i-1,q)$ ($0 \leq q \leq j$). We will compute the values $P(k,j)$ ($0 \leq j \leq k$) in $O(k^2 \cdot Op(n))$ time (using the previous algorithm). Now, we will define the $(k+1) \cdot (k+1)$ matrix $T$ (with rows and columns indexed from $0$ to $k$), where $T(j,q)=1$, if $q \leq j$, and $0$, otherwise. We denote by $PC(i)$ the $(k+1)$-element column vector, where $PC(i)(j)=P(i,j)$. We have $PC(i)=T \cdot PC(i-1)$, for $i \geq k$. Thus, we have $PC(k+1)=T \cdot PC(k)$, $PC(k+2)=T \cdot PC(k+1)=T^2 \cdot PC(k)$, and, in the general case, $PC(i>k)=T^{i-k} \cdot PC(k)$. We can raise the matrix $T$ at any power $p$ in time $O(M(k+1) \cdot \log(p))$, where $M(r)$ is the best time complexity for multiplying two $r$ by $r$ matrices. We can easily have $M(r)=O(r^3)$, but we can also have $M(r)=O(r^{2.807})$. Then, the value of $P(n,k)$ is $PC(n)(k)$. In order to raise a square $d$-by-$d$ matrix $A$ to the $p^{th}$ power, we proceed as follows. We initialize the result matrix $Res=I_k$. Then, we consider the binary representation of $p$ : $b(BMAX), b(BMAX-1), \ldots, b(0)$ (where $b(j)$ can be $0$ or $1$ and $p$=the sum of the values $b(j) \cdot 2^j$, with $0 \leq j \leq BMAX$; $BMAX$ is the index of the most significant bit). We then consider the bits $j$ in reverse order (from $BMAX$ down to $0$). For each such bit $j$, we first set $Res=Res^2$; then, if $b(j)=1$, we further set $Res=Res \cdot A$.

## 7. RELATED WORK

Many multicast tree construction and maintenance techniques have been proposed in the literature. In (Tran et al., 2003), the authors present ZIGZAG, a multicast tree architecture in which every peer has $O(K^2)$ degree and the diameter is $O(\log_K(N))$. ZIGZAG has a hierarchical structure and whenever a new peer joins the tree, it contacts the root first, which redirects it to another peer, and so on. In comparison, our method can provide $O(\log_K(N))$ diameter with $O(K)$ node degree and the tree structure is not hierarchical – any node can be contacted when a new peer wants to join the tree; the non-hierarchical structure avoids the upper level congestion which may occur in ZIGZAG. Single- and multiple-tree approaches based on structured peer-to-peer systems were presented in (Rowstron et al., 2001) and (Castro, 2003). However, some of these systems may also maintain other connections except for those used by the multicast tree. Collaborative multiple multicast tree approaches were presented in (Padmanabhan et al., 2002) and (Venkataraman et al., 2006). In (Cohen and Kaempfer, 2001), the authors consider several optimization objectives for constructing a multicast tree (e.g. a maximum bottleneck multicast tree). Other concerns regarding multicast trees are anonymity (Xiao et al., 2006) and transfer reliability (Andreica and Tapus, 2008a). In (Lin et al., 2006), a problem which is similar to the vertex covering by multicast groups was considered. Maximum profit scheduling problems with particular conflict graphs were considered in (Andreica and Tapus, 2008b).

## 8. CONCLUSIONS AND FUTURE WORK

In the first part of this paper we presented a scalable peer-to-peer multicast tree architecture with bounded degree and small diameter which supports dynamic node arrivals and departures. The architecture converges to a theoretically optimal structure at low node arrival and departure rates. The system was analyzed both from a theoretical and a practical point of view (using simulations) and the results are very good. In the second part of the paper we considered several offline data distribution optimization problems (e.g. covering tree vertices by multicast groups, maximum profit scheduling using conflict graphs, and so on), for which we presented new and efficient algorithmic solutions.